\documentclass[10pt]{iopart}

\usepackage{amsbsy,amssymb,natbib}
\usepackage{citesort}

\usepackage{hyperref}
\usepackage{mathrsfs}
\usepackage{latexsym}
\usepackage{amsbsy}
\usepackage{amssymb}
\usepackage{epsfig}
\usepackage{graphicx}
\usepackage{graphicx}
\usepackage{bm}
\usepackage{natbib}
\usepackage[T1]{fontenc}

\begin{document}

\title[The effect of pressure gradients on luminosity distance -- redshift relations]{The effect of pressure gradients on luminosity distance -- redshift relations\footnote{Research undertaken as part of the Commonwealth Cosmology Initiative (CCI: www.thecci.org) an international collaboration supported by the Australian Research Council.}}

\author{Paul D. Lasky$^{1,2}$, Krzysztof Bolejko$^{3,4}$}
\address{$^{1}$Centre for Astrophysics and Supercomputing, Swinburne University, Hawthorn, Australia}
\address{$^{2}$Theoretical Astrophysics, Eberhard-Karls University of T\"ubingen, T\"ubingen 72076, Germany}
\address{$^{3}$Department of Mathematics and Applied Mathematics, University of Cape Town, South Africa}
\address{$^{4}$Nicolaus Copernicus Astronomical Center, Bartycka 18, 00-716 Warsaw, Poland}
\ead{lasky@tat.physik.uni-tuebingen.de, bolejko@camk.edu.pl}

%\label{firstpage}

		\begin{abstract}
Inhomogeneous cosmological models have had significant success in explaining cosmological observations without the need for dark energy.  Generally, these models imply inhomogeneous matter distributions alter the observable relations that are taken for granted when assuming the Universe evolves according to the standard Friedmann equations.  Moreover, it has recently been shown that both inhomogeneous matter and pressure distributions are required in both early and late stages of cosmological evolution.  These associated pressure gradients are required in the early Universe to sufficiently describe void formation, whilst late-stage pressure gradients stop the appearance of anomalous singularities.  In this paper we investigate the effect of pressure gradients on cosmological observations by deriving the luminosity distance -- redshift relations in spherically symmetric, inhomogeneous spacetimes endowed with a perfect fluid.   By applying this to a specific example for the energy density distribution and using various equations of state, we are able to explicitly show that pressure gradients may have a non-negligble effect on cosmological observations.  In particular, we show that a non-zero pressure gradient can imply significantly different residual Hubble diagrams for $z\lesssim1$ compared to when the pressure is ignored.  This paper therefore highlights the need to properly consider pressure gradients when interpreting cosmological observations.
		\end{abstract}
		
%\begin{keywords}
%cosmology: theory --- cosmology: large-scale structure of Universe
%\end{keywords}
% 98.62.Py	Distances, redshifts, radial velocities; spatial distribution of galaxies
% 98.80.Jk	Mathematical and relativistic aspects of cosmology
% 95.36.+x	Dark energy
% 95.30.Sf	Relativity and gravitation
\maketitle

\section{Introduction}
The idea that we live in a Universe dominated by dark energy is based on a model dependent interpretation of cosmological observations.  A key assumption invoked in these models is that, throughout its evolution, the Universe is both homogenous and isotropic on all scales and is therefore adequately described by a Friedmann-Lema\^itre-Robertson-Walker (FLRW) spacetime.  However, there now exists a wealth of literature that relaxes the homogeneity assumption to explain key cosmological observations without the requirement of dark energy.  In particular, numerous authors have shown that the dimming of supernovae type Ia (SN Ia) may not be due to an accelerated expansion of the Universe, but a change in the predicted luminosity distance -- redshift relation \citep{dabrowski98, pascualsanchez99, celerier00, tomita00,tomita01a,tomita01b,iguchi02, 
godlowski04, alnes06, chung06, alnes07, enqvist07, alexander07, bolejko08a, garciabellido08, garciabellido08b,zibin08,yoo08,enqvist08, bolejko09,celerier09}.  Such inhomogeneous cosmological models have since been used to describe much more than just the dimming of SN Ia data.  Indeed the implementation of such models may suggest we live next to a Gpc scale void \cite[for a short review see][]{ellis08}.  Alternatively, it has recently been shown that inhomogeneous cosmologies utilizing the pressure-free Lema\^itre-Tolman models can also reproduce the SN Ia results by including a ``giant hump'' at the present time \citep{celerier09}.  Moreover, numerous future probes have been suggested based on Baryon Acoustic Oscillations \citep{bolejko09,garciabellido09}, SN Ia measurements at higher redshifts \citep{clifton08}, time-drift of cosmological redshifts \citep{uzan08}, spectral distortions of the cosmic microwave background power spectrum \citep{caldwell08} the kinematic Sunyaev-Zel'dovich effect \citep{garciabellido08b}, the shape of the distance -- redshift relation \citep{clarkson08,february09} and even the cosmic neutrino background \citep{jia08} which may definitively determine the correct model of the Universe.  

There are many different approaches to implement inhomogeneous cosmological models including spatial averaging, pertubative methods and the use of exact solutions of the Einstein field equations \cite[see][for a review of these methods]{celerier07}.  The exact solution method is considered to be the ``\ldots most straghtforward and devoid of theoretical pitfalls'' \citep{celerier07} as they can be used to represent both strong and weak inhomogeneities.  However, these models are difficult to implement due to the limited type of known solutions with either complicated symmetries or matter distributions.  As such, a majority of work to date has utilized the Lema\^itre-Tolman class of solutions which describe a spherically symmetric spacetime endowed with an inhomogeneous distribution of dust \citep{lemaitre33,tolman34}.  Generalizations include the use of Stephani spacetimes \citep{clarkson99,godlowski04,stelmach06,dabrowski07}, Szafron solutions \citep{moffat06} as well as various swiss-cheese models using Lema\^itre-Tolman solutions \cite[for example][]{brouzakis07,marra07, alexander07,brouzakis08,biswas08,kolb09} and Szekeres solutions \citep{bolejko08e} embedded in FLRW spacetimes.  Each of these models have been utilized because exact solutions describing the evolution of the spacetime is known, however this comes at a cost as it implies some sense of physical realism is relaxed.  For example, the Stephani class of solutions have inhomogeneous pressure distributions but the density is only a function of the temporal coordinate, whereas the Lema\^itre-Tolman spacetimes have an inhomogeneous density but zero pressure.  \citet{bolejko06} however, has shown that the formation of observed voids \citep{hoyle04} may be highly dependent on inhomogeneous radiation that is non-zero at the surface of last scattering and soon thereafter.  This was deduced by studying the spherically symmetric Einstein field equations endowed with a perfect fluid with a radiative equation of state.  \citet{bolejko08c} further studied these models to allay previous concerns \citep{hellaby85,bolejko05} that anomalous singularities, known as shell crossing singularities, would occur throughout the evolution of these inhomogeneous spacetimes.  In fact, whilst shell crossing singularities may still occur in Lema\^itre-Tolman spacetimes, the incorporation of non-zero pressure gradients delays the onset of singularity formation until considerably after structure formation has taken place.  This implies that pressure gradients also play an important role in the late term evolution of the Universe \citep{bolejko08c}. For reviews of inhomogeneous cosmological models the reader is referred to \citet{Krasinski97,celerier07,hellaby09}.

The aforermentioned cosmological models attempt to explain observations of the late term evolution of the Universe through inhomogeneities in the baryonic matter.  However, there exists many alternative approaches which also offer promising insights.  In particular, if dark energy exists and is not a cosmological constant but some kind of dynamical field or negative pressure fluid, then it must posses inhomogeneities due to gravitational interactions with both itself and with clumps of baryonic and dark matter.  The effect of these inhomogeneities on structure formation can be non-neglible \cite[e.g.][]{basilakos03,linder03,linder05,kuhlen05,abramo07,abramo08,mota08,avelino08} and possibly requires full non-linear evolution for a complete description \citep{abramo08,abramo09}.  Therefore, in these models the geometry of the spacetime is necessarily inhomogeneous, implying the non-linearities in the Einstein field equations will play a pivotal role, in part due to the non-zero pressure gradients of the inhomogeneous dark energy.

Non-zero pressure gradients enter the Einstein field equations through the energy-momentum tensor, and hence effect the overall geometry of the spacetime.  This therefore effects the way in which we interpret various cosmological observations by changing the overall luminosity distance -- redshift relation.  The purpose of this article is therefore to calculate the luminosity distance -- redshift relation for spherically symmetric, perfect fluid spacetimes where the pressure and energy-density are allowed to vary both in the temporal and radial direction\footnote{Such spherically symmetric, perfect fluid spacetimes were first considered by \cite{lemaitre33} who even allowed for anisotropic pressure distributions.  Later, \citet{podurets64,podurets64a} and \citet{misner64} considered the isotropic pressure case, and the equations have since been accredited in the literature to Misner and Sharp.  However, throughout the present article we refer to these spacetimes as ``Lema\^itre'' models to give the appropriate credit.}.  These models generalize those of Lema\^itre-Tolman spacetimes, which can be recovered by setting the equation of state such that the pressure vanishes.  Moreover, these spacetimes can be applied to dynamical dark energy models by choosing an equation of state where the pressure is negative.  

This paper is set out as follows: In section \ref{spacetime} we show the general form of the sphericaly symmetric, perfect fluid spacetimes, including relating the metric coefficients back to familiar cosmological entities such as the local Hubble rate and the critical densities.  In section \ref{redshift} we derive the overall redshift relation for our spherically symmetric perfect fluid with an observer situated at the coordinate origin (i.e. $r=0$), also the luminosity distance in terms of the metric coefficients, hence arriving at the luminosity distance -- redshift relation.  In section \ref{example} we show an example of the change in luminosity distance -- redshift relation for a specific example of an inhomogeneous cosmological model with differing equations of state, and we provide concluding remarks in section \ref{conc}.  Throughout the paper we use coordinates in which the speed of light is unity unless otherwise explicitly stated.  Moreover, greek indices range over $0\ldots3$.

\section{Spherically Symmetric Perfect Fluid Spacetimes}\label{spacetime}
The evolution of a spherically symmetric perfect fluid was first studied by \cite{lemaitre33}. The metric in comoving coordinates can be expressed as
\begin{eqnarray}
	ds^{2} = -\alpha^{2} dt^{2}+\frac{R'^{2}}{1+E}dr^{2}+R^{2}d\Omega^{2}.\label{pfle}
\end{eqnarray}
Here, $\alpha(t,r)>0$ is the lapse function, $R(t,r)$ is the area-radius coordinate and $E(t,r)>-1$ is the curvature parameter.  Moreover, the two-spheres are denoted by $d\Omega^{2}:=d\theta^{2}+\sin^{2}\theta d\phi^{2}$ and a prime denotes differentiation with respect to the radial coordinate, $r$. 

When the pressure and density are functions of the temporal coordinate only, the lapse function can be set to unity without loss of generality, and 
\begin{eqnarray}
	R\rightarrow a(t)r\qquad{\rm and}\qquad E\rightarrow-kr^{2}.
\end{eqnarray}
This is then the standard FLRW spacetime, where $k=0,\,\pm1$ is the curvature index and $a(t)$ is the scale factor.  

A perfect fluid stress-energy tensor takes the form
\begin{eqnarray}
	T_{\mu\nu}=\left(\rho+P\right)u_{\mu}u_{\mu}+Pg_{\mu\nu},
\end{eqnarray}
where $u^{\mu}$ is the four-velocity of the fluid, which is given in coordinate form by $u_{\mu}=-\alpha{\delta_{\mu}}^{t}$.  

%%%%%%%%%%%%%%%%%%%%%%%%%%%%%%%%%%%%%%%%%%%%%%%%%%%%%%%%
We express the Einstein field equations as $G_{\mu\nu}=\kappa T_{\mu\nu}-\Lambda g_{\mu\nu}$, where $\Lambda$ is the cosmological constant and $\kappa=8\pi G$.  The ${G_{t}}^{r}$ component of the Einstein field equations is
\begin{eqnarray}
	\frac{\dot{E}}{2\left(1+E\right)}-\frac{\alpha'\dot{R}}{\alpha R'}=0,\label{Edot}
\end{eqnarray}
where an overdot denotes partial derivative with respect to temporal  coordinate, i.e. $\dot{E} = \partial E / \partial t$,
and a prime denotes partial derivative with respect to radial coordinate, i.e. $\alpha' = \partial \alpha / \partial r$.
In the dust limit, the second term in this equation vanishes, implying $E=E(r)$.  The above equation is therefore an extra equation that must be satisfied from both the dust and the homogeneous cases.  The remaining Einstein field equations can be expressed as
\begin{eqnarray}
	\frac{\dot{R}^{2}-E\alpha^{2}}{\alpha^{2}R^{2}}+\frac{2\dot{R}\dot{R}'-E'\alpha^{2}}{\alpha^{2}R'R}-\frac{\dot{R}\dot{E}}{\alpha^{2}R\left(1+E\right)}&=\kappa\rho+\Lambda,\label{one}\\
	\dot{R}^{2}+2R\ddot{R}-E\alpha^{2}-\frac{2R\dot{R}\dot{\alpha}}{\alpha}-\frac{2R\alpha\alpha'\left(1+E\right)}{R'}&=-R^{2}\alpha^{2}\left(\kappa P-\Lambda\right).\label{two}
\end{eqnarray}
It is again noted that, in the dust case, the final term on the left hand side of (\ref{one}), and the last two terms on the left hand side of (\ref{two}) vanish, and the equations reduce to those of the dust version given by \cite{enqvist08}, and indeed without these terms the equations are remarkably similar to the set of equations for homogeneous FLRW spacetimes.  Both the final terms on the left hand sides of equations (\ref{one}) and (\ref{two}) are associated with the pressure gradient through equations (\ref{Euler}) and (\ref{Edot}).  

The first integral of equation (\ref{two}) can be found to be
\begin{eqnarray}
	\frac{\dot{R}^{2}}{\alpha^{2}R^{2}}=\frac{F}{R^{3}}+\frac{\Lambda}{3}+\frac{E}{R^{2}},\label{Fried}
\end{eqnarray}
where $F(t,r)$ is twice the Misner-Sharp mass \citep{misner64}.  This equation is now expressed analogously to that of the FLRW and pressure-free Lema\^itre-Tolman cases \cite[for example see][]{enqvist07}.  Indeed, the only differences, aside from the additional functional dependencies, is the presence of the lapse function, $\alpha$, which we will later show becomes incorporated into the definition of the local Hubble rate.

Putting this back through equations (\ref{two}) and (\ref{one}) respectively gives
\begin{eqnarray}
	\dot{F}&=-\kappa P\dot{R}R^{2},\label{Fdot}\\
	F'&=\kappa\rho R'R^{2}.\label{Fdash}
\end{eqnarray}
When $F'$ remains non-zero and $R'$ or $R$ tend to zero, equation (\ref{Fdash}) implies the density diverges, which is interpreted as shell crossing and shell focussing singularities respectively \cite[for a discussion of shell crossing singularities in a cosmological context see][and references therein]{bolejko08c}.

%%%%%%%%%%%%%%%%%%%%%%%%%%%%%%%%%%%%%%%%%%%%%%%%%%%%%%%%

The radial component of the conservation of energy momentum, ${T^{\alpha}}_{\mu;\alpha}=0$, implies
\begin{eqnarray}
	\left(\rho+P\right)\frac{\alpha'}{\alpha}=-P'.\label{Euler}
\end{eqnarray}
This equation says that the pressure gradient being non-zero, i.e. $P'\neq0$, implies that the observer is non-geodesic.  That is, the presence of a pressure gradient pushes the observer off the freely falling geodesics.  If there is zero pressure gradient, then the lapse function is also a function of only the temporal coordinate, which can then be rescaled such that $\alpha=1$.  This implies that an observer in a Universe with zero pressure gradient is freely falling along with the fluid.  Moreover, equations (\ref{Edot}) and (\ref{Euler}) imply that the temporal evolution of the curvature parameter, $E$, is governed by the pressure gradient.  That is, if the pressure distribution were homogeneous (even if it were non-zero), then the curvature parameter would not evolve in time.  

The temporal component of the conservation equations can be manipulated using equation (\ref{Euler}) and (\ref{Edot}) to give
\begin{eqnarray}
	\dot{\rho}+P'\frac{\dot{R}}{R'}=-\left(\rho+P\right)\left(\frac{\dot{R}'}{R'}+\frac{2\dot{R}}{R}\right).
\end{eqnarray}
This equation has been expressed such that the explicit dependence on the pressure gradient can again be seen.  In the Lema\^itre-Tolman case, i.e. where $P=0$, and the FLRW case, i.e. where $P'=0$, the second term on the left hand side of the above equation vanishes.  Moreover, in the FLRW case, the right hand side can be shown to be $-3\left(\rho+P\right)\dot{R}/R$.

%-----------------------------------------

We now define the local Hubble rate to be
\begin{eqnarray}
	H(t,r):=\frac{\dot{R}}{\alpha R}.\label{localHubble}
\end{eqnarray}
The additional dependence on the lapse function, $\alpha$, in this definition comes from equation (\ref{Fried}), and is attributed to the observer no longer traveling along geodesics due to the introduction of a pressure gradient.  Therefore, different observer will measure a different local Hubble expansion rate\footnote{This is also an underlying premise in the Fractal Bubble Cosmological models of Wiltshire and collaborators \citep{wiltshire07a,wiltshire07b,wiltshire08}, although see \citet{kwan09} for a recent critique of these models.}.

We define the current local Hubble rate, $H_{0}(r):=H\left(t_{0},r\right)$, defined on some spacelike hypersurface, given by $t=t_{0}$, and also the current area-radius coordinate, $R_{0}(r):=R\left(t_{0},r\right)$.  In FLRW cosmology (with non-zero pressure contribution), the Hubble rate is given in terms of the critical densities according to
\begin{eqnarray}
	\widetilde{H}^{2}=\widetilde{H}_{0}^{2}\left[\widetilde{\Omega}_{M}\left(\frac{R}{R_0}\right)^{-3(1+\tilde{\omega})}+
\widetilde{\Omega}_{c}\left(\frac{R}{R_0}\right)^{-2} +
\widetilde{\Omega}_{\Lambda}
\right],\label{Hublaw}
\end{eqnarray}
where $\widetilde{\Omega}_{c}:=1-\widetilde{\Omega}_{\Lambda}-\widetilde{\Omega}_{M}$ and tildes are used to denote quantities in the FLRW model.  Analogously, from equation (\ref{Fried}) we make the following definitions for the various critical densities in terms of the metric coefficients in the Lema\^itre models:
\begin{eqnarray}
\Omega_{M} (t_0,r) &= \frac{F_0}{H_0^2 R_0^3}, \label{Feqn}\\
\Omega_{\Lambda} (t_0,r) &= \frac{\Lambda}{3H_0^2},\label{Leqn}\\
\Omega_{c} (t_0,r) &= \frac{E_0}{H_0^2 R_0^2} 
= 1-\Omega_{\Lambda}-\Omega_{M}\label{Eeqn}.\
\end{eqnarray}
This implies equation (\ref{Fried}) is now expressed as 
\begin{eqnarray}\label{InhHL}
H^{2}(t,r)=&H_{0}^{2}(r)\left[\Omega_{M} \left(\frac{R}{R_0}\right)^{-3(1+f)}+\Omega_{c} \left(\frac{R}{R_0}\right)^{-2(1+e)} 
+\Omega_{\Lambda}(r)\right],
\end{eqnarray}
where
\begin{eqnarray}
\left(\frac{R}{R_0}\right)^{-3f} = \frac{F}{F_0}\qquad{\rm and}\qquad\left(\frac{R}{R_0}\right)^{-2e} = \frac{E}{E_0}.
\end{eqnarray}
Thus, in analogy to FLRW spacetimes we see three qualitative differences.
Firstly, we see that the $\Omega$'s are no longer constant parameters
but they depend on the radial coordinate, $r$.  Secondly, $f$ in (\ref{InhHL}) is no longer
the equation of state as $\tilde{\omega}$ is in (\ref{Hublaw}).
Lastly, and most importantly, the 
function $E$ depends on time (in FLRW models $E/r^2$ is constant), it
is no longer constant and its evolution affects the expansion rate.
This is an important observation since, for
example evolving homogeneous dark energy with a given equation of state,
$w(z)$, will generally behave differently than an inhomogeneous dark energy fluid.

In the Lema\^itre-Tolman dust case equation (\ref{Hublaw}) is separable and one can therefore find an implicit solution for $\dot{R}$ \cite[see equation 2.22 of ][]{enqvist08}.  However, this is not the case when there exists a pressure gradient, implying we require numerical rather than analytic solutions.

\section{Distance -- Redshift Relations}\label{redshift}
Herein we consider an observer at the origin of the coordinate system.  The spherically symmetric nature of the spacetime implies all observed photons coming from our past null cone have come along radial null geodesics.  Therefore, setting $d\theta=d\phi=ds^{2}=0$ in the line element (\ref{pfle}) implies
\begin{eqnarray}
	\frac{dt}{du}=-\frac{dr}{du}\frac{R'}{\alpha\sqrt{1+E}}:=-\frac{dr}{du}\chi(t,r),\label{par}
\end{eqnarray}
where $u$ parametrizes the geodesics and $\chi$ has been defined for simplicity in the following expressions.  Now we consider two light rays separated by a small distance $\lambda(u)$, such that the first and second light rays are expressed as a parametrization of $u$ as $t_{1}(u)=t(u)$ and $t_{2}(u)=t(u)+\lambda(u)$ respectively.  Differentiating these with respect to the parametrization, and substituting equation (\ref{par}) gives
\begin{eqnarray}
	\frac{dt_{1}}{du}&=-\frac{dr}{du}\chi(t,r),\label{t1}\\
	\frac{dt_{2}}{du}&=-\frac{dr}{du}\chi(t,r)+\frac{d\lambda}{du}.\label{t21}
\end{eqnarray}
The equation for $t_{2}$ can also be calculated using a Taylor series approximation such that
\begin{eqnarray}
	\frac{dt_{2}}{du}&=-\frac{dr}{du}\chi\left(t+\lambda,r\right)\nonumber\\
		&=-\frac{dr}{du}\left[\chi\left(t,r\right)+\frac{\partial\chi\left(t,r\right)}{\partial t}\lambda\right],\label{t22}
\end{eqnarray}
where terms non-linear in $\lambda$ have been neglected.  This is a valid assumption as the separation of the two light rays is assumed to be much less than the distance over which we are considering the propagation of the light rays.  Evaluating the right hand side of equations (\ref{t21}) and (\ref{t22}) gives a differential equation for $\lambda(u)$,
\begin{eqnarray}
	\frac{d\lambda}{du}=-\frac{dr}{du}\frac{\partial\chi}{\partial t}\lambda.
\end{eqnarray}
Finally, the redshift is defined as
\begin{eqnarray}
	z(u):=\frac{\lambda(0)-\lambda(u)}{\lambda(u)},
\end{eqnarray}	
Differentiating this, and using the above relations implies
\begin{eqnarray}
	\frac{dz}{du}&=\frac{dr}{du}\frac{\partial\chi}{\partial t}\left(1+z\right)\nonumber\\
			&=\frac{dr}{du}\frac{\left(1+z\right)R'}{\alpha\sqrt{1+E}}\left(\frac{\dot{R}'}{R'}-\frac{\dot{\alpha}}{\alpha}-\frac{\dot{E}}{2\sqrt{1+E}}\right).\label{that}
\end{eqnarray}
Finally, utilizing equation (\ref{par}) and (\ref{that}), one can derive a pair of differential equations which give the coordinates in terms of the observable redshifts\footnote{ Here we have implicitly assumed that the redshift is a single-valued function of the geodesic parametrization, $u$.  In Lema\^itre-Tolman dust spacetimes this has been shown to not necessarily be the case depending on initial density configurations \citep{mustapha98}, and we note that it requires verification for the Lema\^itre spacetimes with non-zero pressure distributions.};
\begin{eqnarray}
	\frac{dt}{dz}&=\frac{-\chi}{\dot{\chi}\left(1+z\right)},\label{tz}\\
	\frac{dr}{dz}&=\frac{1}{\dot{\chi}\left(1+z\right)},\label{rz}
\end{eqnarray}
where 
\begin{eqnarray}
	\dot{\chi}=\frac{R'}{\alpha\sqrt{1+E}}\left(\frac{\dot{R}'}{R'}-\frac{\dot{\alpha}}{\alpha}-\frac{\dot{E}}{2\sqrt{1+E}}\right).
\end{eqnarray}
The $\dot{E}$ term is related through equation (\ref{Eeqn}) to the temporal evolution of the radiation and matter densities.  That is, the way one views the redshift of an object depends on the evolutionary history of terms such as $\Omega_{R}$ and $\Omega_{M}$.  

We now derive relations for the distance measurements in terms of the spacetime variables.  The area distance, $D_{A}$, is found as the root of the ratio of the angle subtended by null geodesics diverging from the observer to the cross-sectional area of this bundle. For a general metric in spherical coordinates, null geodescis propagating along the constant $\theta$ and $\phi$ direction, the angular distance is \citep{ellis85,stoeger92}
\begin{eqnarray}
	D_{A}^2= \frac{\sqrt{|\det j_{ab}|}}{\sin \theta} = R^2(t,r),
\end{eqnarray}
where $j_{ab}$ is the $\theta-\phi$ part of the metric.  The luminosity distance, $D_{L}$, and the area distance are related by ${D_{L}}^{2}={D_{A}}^{2}\left(1+z\right)^{4}$ \citep{etherington33} implying in the spherically symmetric perfect fluid spacetime,
\begin{eqnarray}
	D_{L}=R\left(1+z\right)^{2}.\label{lumdist}
\end{eqnarray}
Equations (\ref{tz}), (\ref{rz}) and (\ref{lumdist}) now provide the necessary equations to determine the luminosity distance -- redshift relation for specific matter distributions in a Lema\^itre spacetime with a central observer.

Finally, the luminosity distance allows us to work out the apparent bolometric magnitude, $m$, of a standard candle from the absolute bolometric magnitude, $M$, at a given redshift.  This is expressed in Mpc as
\begin{eqnarray}
	m=M+5\log D_{L}\left(z\right)+25
\end{eqnarray}
where $D_{L}$ depends on all $\Omega_{a}$ terms and the local Hubble rate.  This expression is required in section \ref{example}  below for calculating the residual Hubble diagrams.

\section{Inhomogeneous Example}\label{example}
We now provide an example of a calculation for an inhomogeneous cosmological model.  To determine the effect of the pressure gradient on the luminosity distance -- redshift relation, we concentrate on two models: the pressure-free Lema\^itre-Tolman model and the Lema\^itre model that we have been dealing with hitherto.  We specify both models with the same initial data at the current insant.  Moreover, the background model is chosen to be an open Friedmann model with 
$\Omega_{mat} = 0.3$, $\Omega_{\Lambda} =0$ and $H_0  = 67.2$ km s$^{-1}$ Mpc$^{-1}$.
The radial coordinate is defined as the areal distance, $R$, at the current instant:
$\tilde{r}:= R(t_0,r)$. However, for clarity in further use,  
the tilde sign is omitted and the new radial coordinate is referred to as $r$.
The pressure-free Lema\^itre-Tolman model is of a Gpc-scale void model,
with ${E}_0$ and ${F}_0$ [note from equations (\ref{Euler}), (\ref{Edot}), and (\ref{Fdot}) that for pressureless matter these functions depend on $r$ only] calculated in the following way.
The ${F}_0$ function is found from (\ref{Fdash}) with $\rho$ given by:
\begin{eqnarray}
	\rho(t_0,r) = \rho_b \left[ 1 + \delta - \delta \exp \left( - \frac{r^2}{\sigma^2} \right) \right],
\label{rhofl}
\end{eqnarray}
where $\rho_b = 0.3 \times  (3H_0^2)/(8\pi G)$, $\delta = 1.5$, and $\sigma= 1$ Gpc.
The function ${E}_0$ is evaluated by integrating (\ref{Fried}) and assuming
the the cosmological and integration constants are zero:
\begin{eqnarray}\label{EinLT}
	\int\limits_0^R\frac{d r}{\sqrt{{E}_0 + {F}_0/r}} = c t_0.
\end{eqnarray}
Initial data for the Lema\^itre model is $E(t_0,r)={E}_0$ and $F(t_0,r) = {F}_0$.
In addition we assumed the following equation of state:
\begin{eqnarray}\label{eos}
	p = K \rho^n,
\end{eqnarray}
where $K$ and $n$ are constant.
We then use eqations (\ref{tz})-(\ref{rz}) to find $t$ and $r$ along the past null cone
and simultaneously solve (\ref{Edot}), (\ref{Fried}), and (\ref{Fdot}) to find the 
evolution of the system.

\begin{figure}
\begin{center}
\includegraphics[scale=0.7]{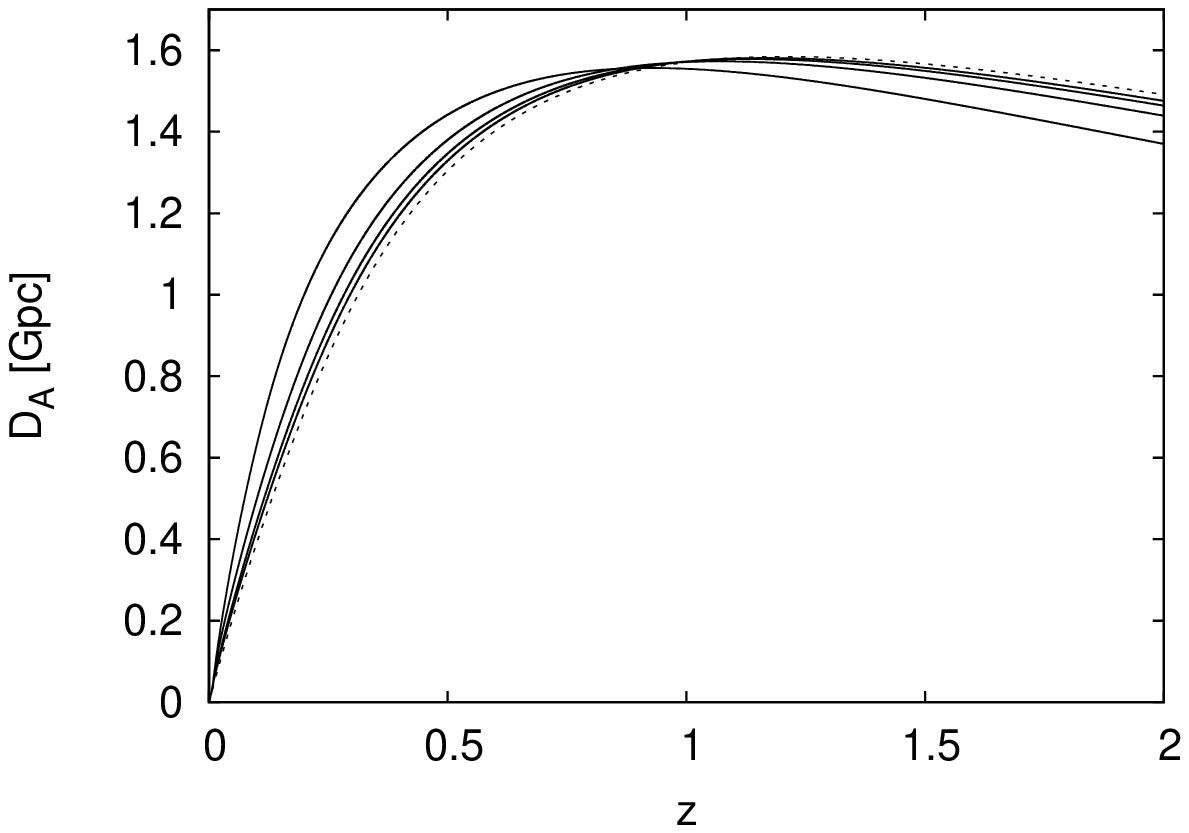}
\caption{Angular diameter distance -- redshift diagram for models with equation of state $P=K\rho$.  The dashed curve has $K=0$, and the solid curves from the top to the bottom (at the maximum) have $K=0.0005$, $0.001$, $0.002$ and $0.005$.  } 
\label{ADDvarK}
\end{center}
\end{figure}

\begin{figure}
\begin{center}
\includegraphics[scale=0.7]{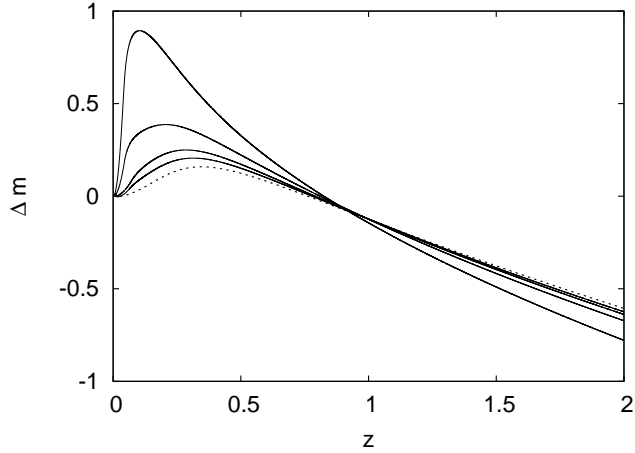}
\caption{Residual Hubble diagram for models with equation of state $P=K\rho$.  The dashed curve has $K=0$, and the solid curves from the top to the bottom (at the maximum) have $K=0.005$, $0.002$, $0.001$ and $0.0005$.} 
\label{HDvarK}
\end{center}
\end{figure}

\begin{figure}
\begin{center}
\includegraphics[scale=0.7]{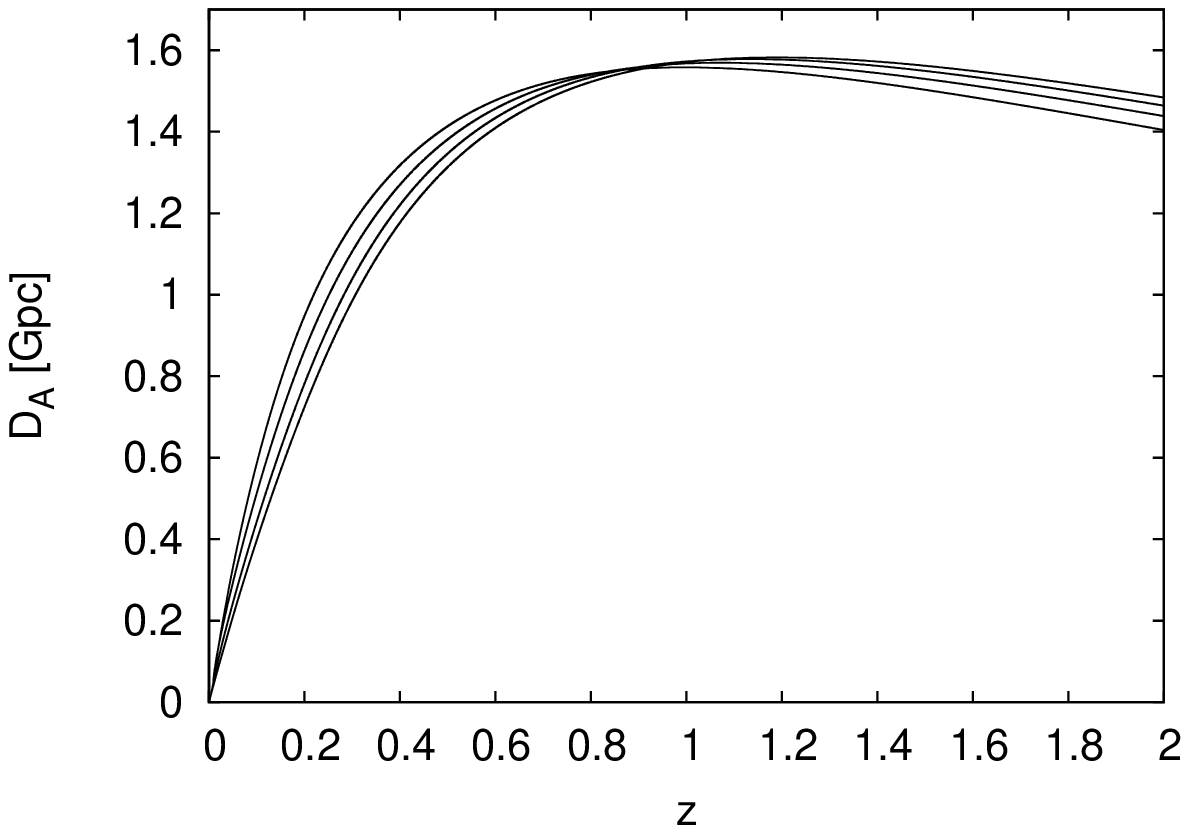}
\caption{Angular diameter distance as a function of redshift for models with equation of state $P=10^{-3}\rho^{n}$.  The curves from top to bottom (at the maximum) have $n=1.05$, $1.00$, $0.97$ and $0.95$. }
\label{ADDvarn}
\end{center}
\end{figure}

\begin{figure}
\begin{center}
\includegraphics[scale=0.7]{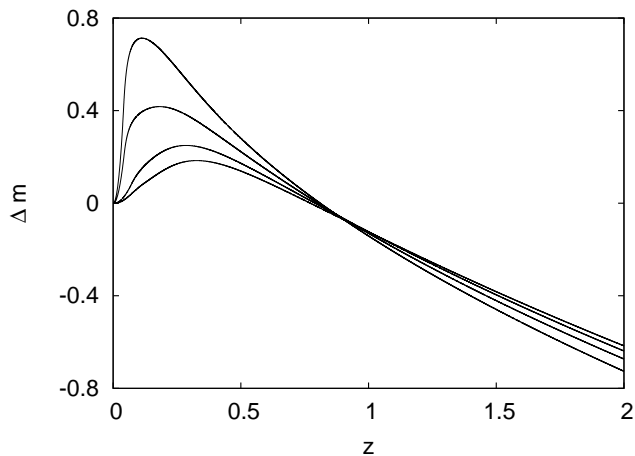}
\caption{Residual Hubble diagram for models with equation of state $P=10^{-3}\rho^{n}$.  The curves from top to bottom (at the maximum) have $n=0.95$, $0.97$, $1.00$ and $1.05$.} 
\label{HDvarn}
\end{center}
\end{figure}

Our results are presented in figures \ref{ADDvarK}--\ref{HDvarn}.  Figures \ref{ADDvarK} and \ref{HDvarK} present the calculated angular diameter distance -- redshift and residual Hubble diagram respectively for different values of the equation of state parameter $K$.  Figures \ref{ADDvarn} and \ref{HDvarn} are respectively the same figures where we now vary the parameter $n$.  

It is apparent that for a linear equation of state, an increase in the proportionality constant, $K$, results in increasingly larger differences in the distance measurements (figures \ref{ADDvarK} and \ref{HDvarK}).  For low redshifts, $z\lesssim1$, the pressure-free Lema\^itre-Tolman model has smaller values of the distance at equal redshifts.  Moreover as $K$ increases, so too does the value of the distance.  This feature is better depicted in the residual Hubble diagram, figure \ref{HDvarK}, where the variation in magnitude as a function of redshift is large at small redshifts but becomes less distinguishable for $z\gtrsim1$.  Deviations from a linear equation of state are depicted in figures \ref{ADDvarn} and \ref{HDvarn}.  In these figures we have only varied $n$ between $1.05$ and $0.95$, thereby indicating the sensitive nature of this parameter on the distance measurements.  This is due to small variations in $n$ implying large variations in the magnitude of the pressure gradient with respect to the energy-density.  Again, the residual Hubble diagram highlights the importance of this parameter at low redshifts, $z\lesssim1$, where a variation of $n$ by $0.1$ can cause $\Delta m$ to vary by up to $0.6$.

Another interesting feature is that the position of the maximum of the angular diameter distance shifts to lower redshifts as the amplitude of the pressure gradient increases. As recently shown by \citet{AH09} the position of the apparent horizon 
(${\rm d} D_A /{\rm d} z = 0$) for non-zero pressure models is given by the same relation as for the dust models, i.e. $R=2M$.  However, since for $p'\ne0$ the function $M$ depends on time and thus this relation is satisfied by a different pair of $t$ and $r$ than in the dust models.

As can be seen from figures \ref{ADDvarK}--\ref{HDvarn}, the incorporation of pressure gradients in cosmological models can significantly alter the distance -- redshift relations.  For models with a positive pressure gradient, the luminosity distance for a given redshift is larger compared with models for $P=0$.  It is therefore apparent that cosmological models with non-trivial pressure distributions will yield different results than cosmological models that do not consider pressure gradients.

\section{Conclusion}\label{conc}

Whilst a considerable amount of literature has been dedicated to the effect of inhomogeneous cosmological models to explain the observed distance -- redshift relations from SN Ia data \citep{dabrowski98, pascualsanchez99, celerier00, tomita00,tomita01a,tomita01b,iguchi02, 
godlowski04, alnes06, chung06, alnes07, enqvist07, alexander07, bolejko08a, garciabellido08, garciabellido08b,zibin08,yoo08,enqvist08, bolejko09,celerier09}, surprisingly little has been done on the effect of pressure gradients in such models.  The most simple invocation of pressure gradients into inhomogeneous cosmologies is to utilise spherically symmetric Lema\^itre models with an observer placed at the center of symmetry.  In \citet{bolejko08c}, we studied these models to show that anomalous singularities that may occur in the evolution of such {\it dust} spacetimes \citep{hellaby85,bolejko05} are delayed by the inclusion of realistic pressure gradients until significantly after structure formation has taken place.  In the present article we have continued this line of research to derive the distance -- redshift relations in spherically symmetric spacetimes endowed with a perfect fluid.  Moreover, we worked in comoving coordinates with the observer at the center of symmetry.  This work will therefore allow for an interpretation of cosmological data using such models.  

By exploiting specific examples of inhomogeneous matter distributions with various ``simple'' equations of state, we have shown that the introduction of pressure gradients can have a non-negligble effect on observations, particularly for $z\lesssim1$ (figures \ref{ADDvarK}--\ref{HDvarn}).  This implies that one should definitely consider the full effect of pressure gradients in interpreting cosmological data-sets.  However, this is a non-trivial task due to the increased number of free-parameters in the system.  Indeed, if the data can be realistically fit using a void model with $P=0$, then it can equally well be fit by implementing slightly different matter distributions with $P\neq0$.  The problem then becomes one of reverse engineering the correct spacetime geometry from the numerous observations, a task which has been discussed in great detail in \cite{LH07,MH08}.

We conclude by remarking that this work may also have a significant effect on interpretations of models with dark energy which is not a cosmological constant, but instead a dynamical field or negative pressure fluid \cite[e.g.][]{basilakos03,linder03,linder05,kuhlen05,abramo07,abramo08,mota08,avelino08,abramo09}.  These models necessarily have inhomogeneous spacetime geometries, which is equally attributed to ``pressure'' gradients of the dark energy field.  In this way, the distance -- redshift relations that we have derived herein will be equally valid, and can therefore also be used to interpret cosmological data within such models.

\ack
PL was supported under Australian Research Council's Discovery Projects funding scheme (project number DP0665574).

\bibliographystyle{harvard}
\bibliography{pfd_cqg}

\end{document}